\documentclass[iop, apj, revtex4]{emulateapj}
\usepackage{natbib}
\usepackage{url}
\usepackage{amsmath}
\usepackage{appendix}
\usepackage{hyperref}
\begin{document}

\title{Cold Exponential Disks from Interstellar Fountains}

\author{
Curtis Struck\altaffilmark{1}, Bruce G. Elmegreen\altaffilmark{2}}

\altaffiltext{1}{ Dept. of Physics and Astronomy, Iowa State Univ., Ames, IA 50011
USA; curt@iastate.edu}

\altaffiltext{2}{IBM Research Division, T.J. Watson Research Center, 1101
Kitchawan Road, Yorktown Heights, NY 10598, USA; bge@us.ibm.com}

\begin{abstract}
We present the results of a simple numerical model with phenomenological cloud
growth and explosive disruption processes, and with fountain launched ballistic
motions of disrupted cloud fragments out of the disk. These processes generate
an effective scattering of gas elements over much larger distances than
noncircular impulses in the plane, which are quickly damped. The result is
evolution of the global cloud density profile to an exponential form on a roughly Gyr timescale. This is
consistent with our previous results on the effects of star scattering off
massive clumps in young disks, and gas holes in dwarf galaxies.
However, in those cases the scattering processes generated thick, warm/hot
stellar disks. Here we find that the exponential gas disks remain cold.
Star formation in this gas would produce a thin exponential stellar disk.
\end{abstract}


\section{Introduction}
\label{intro}

The origin of the exponential radial profile in galaxy disks is not well
understood. Profiles close to exponential can result from galaxy formation, with
\citep{aumer13,martig14,herpich15,rathaus16} or without \citep{mestel63,freeman70}
specific angular momentum conservation, and two gassy disks can adjust after a
merger to be an exponential \citep{athan16,borlaff14}. However, young galaxies are
observed to be highly irregular \citep{elmegreen05,conselice06} and subsequent
mass accretion can be irregular too \citep[e.g.,][]{ceverino16b}. This makes the
extreme regularity and smoothness of radial profiles in today's old stellar disks
somewhat puzzling. Such regularity seems to require constant re-adjustments to
smooth out environmental perturbations and remake the exponential shape.

We have proposed that stellar scattering from clouds \citep[][hereafter Paper
I]{bour07, es13} and interstellar holes \citep[][hereafter Paper II]{se17} in galaxy disks
can maintain and even make an exponential from some very different initial
structure. Purely random scattering with a slight inward radial bias can make an
exponential as stars migrate around the disk \citep{es16}. Such a bias might arise
from angular momentum perturbations of initially circular orbits with
energy-conserving collisions such as those between stars and massive clouds.
Cloud-like scattering is favored over spiral-arm scattering in dwarf irregular
galaxies which have exponential profiles without spirals or bars
\citep{herrmann13}. Cloud scattering might also be preferred over viscous
evolution \citep{lin87} because dwarfs have little shear.

Stellar scattering from midplane clouds and holes tends to be self-limiting
(Papers I, II). The stars scatter both parallel and perpendicular to the disk and
those with high scale heights do not interact with clouds as much \citep{la84}.
The velocity dispersion can get large, however, and this can be a problem for the
scattering model if a major radial readjustment is necessary before the
exponential forms.

Here we show that exponentials also form in dissipative gas that explodes out of
the disk and falls back down to mix with other disk gas. Examples of observed exponential gas profiles are in e.g., \citet{wong02}, \citet{leroy09}, and \citet{gallagher18}, and for high redshift profiles see \citet{fujimoto18}. For recent models of widespread disk outflows, see e.g., \citealt{ceverino16a, martizzi16}. This dissipative process
does not lead to large velocity dispersions even when the disk is rearranged
completely. A process like this that forms exponentials in gas might be preferred
to purely stellar scattering because the gas is more fundamental than stars in
building a disk. Even in the present universe this process can be efficient, since a significant amount of gas is lofted out of disks with strong star formation; e.g., roughly $10 - 20 \%$ of the total HI mass in some cases (see \citealt{Vargas17} on NGC 4559 and references therein to other HALOGAS studies). Star formation in this reconfigured gas would then produce the stellar
exponential, although with a slightly different scale length in proportional to
the power of the non-linear Kennicutt-Schmidt relation.

In what follows, section \ref{models} describes the basic model, section
\ref{results} presents the results, and section \ref{concl} gives the conclusions.

\section{Models}
\label{models}

The effects of gas mixing on radial profiles is isolated by using an idealized
model of a disk of clouds with a fixed potential for rotational motions and no
other torques that might also re-arrange the disk. The clouds are test particles
initially in circular orbits, and the rotational potential is a power-law in
radius (see Paper I). The acceleration is given by,

\begin{equation}
\label{eqa}
g(r) = \frac{-GM_H}{H^2}
\left(\frac{r + 0.2}{H} \right) ^{-\gamma},
\end{equation}

\noindent where $r$ is the central radius in three dimensions, and the 0.2
term is an arbitrary softening constant. In this potential the disc has a
circular velocity that increases gradually as $R^\frac{1- \gamma}{2}$, where $R$
is the projected radius within the disc. The gravitational mass within a radius
$r$ is $M(r) = M_H (r/H)^{2-\gamma}$.

In the direction perpendicular to the disk, there is a vertical component of the total gravity implicit in  equation (1), plus a vertical gravity from the disk mass itself. To consider the second component, we use an effective gravity in the vertical direction to help hold the disk in place,

\begin{equation}
\label{eqaa}
g_z(r) = \frac{-0.3GM_H}{H^2}
\left( \frac{z}{H} \right).
\end{equation}

\noindent This acceleration is linear in $z$, and with a moderate magnitude,
allows the formation of a  thick disk if the vertical dispersion is high.

We use dimensionless units, i.e, $H = 1,\ T = H/V_H,\ V_H^2 = GM_H/H$, with $GM_H
= 1.0$. We will use two representative scalings for a normal and dwarf type
disk, respectively. For the normal disk these are: $H = 1$ kpc, $V_H = 50$ km
s$^{-1}$, $T = 20$ Myr, and $M_H = 2.9 \times 10^{7}\; M_{\odot}$. Then, the
rotation periods at $R = 2$ and $10$ kpc for a flat rotation curve of magnitude 6 velocity units are $56$ and $303$ Myr, respectively. For
the dwarf disk these are: $H = 0.5$ kpc, $V_H = 25$ km s$^{-1}$, $T = 20$ Myr,
and $M_H = 7.2 \times 10^{7}\;M_{\odot}$. 

The particles are initialized with a mass of one unit, but build up in the merging phase (see below). These mass units are arbitrary, since
there are no gravitational interactions between the clouds. In each computational
cycle, particles with masses of three units or more have a finite probability of
being broken up by feedback effects, thereby ``exploding''. This probability is
set to 0.1 for particles of mass 3.0 units, and it increases by 0.05 for each
increase of particle mass by one unit, up to 1 for masses greater than or equal to
22 units. Other values of these parameters have been tested, and found to not
change the results qualitatively.

The computational cycle has three parts: 1) an explosion phase representing star
formation feedback in the interstellar medium; 2) a phase of particle ballistic
motion in the fixed galaxy potential, and 3) a phase of merging of adjacent
particles to make bigger clouds. In each stage all particles are affected simultaneously.  In the explosion phase, all selected particles of
mass $N$ units are divided into $N$ particles, and each is given
additional random velocity impulses in the x, y, and z directions. The average
magnitude of these velocity increments ranges from 0.3-4.0 code units, where peak
circular velocities are typically 6-10 code units.

In the second phase, we adopt a picture like that of \cite{marasco12} who assumed
that clouds were launched from superbubbles onto nearly ballistic orbits, despite
their interaction with coronal gas. We also neglect this interaction. The
assumption of nearly ballistic orbits allows gas clouds to 
scatter across significant parts of the disk over time. The duration of the ballistic phase
is generally taken equal to 2.0 time units. The exact value is not important as
long as there is enough time for significant (non-circular) motions of the
component particles.

In the merging phase, a grid is imposed on the disk and all particles in each grid
cell are merged as expected from cloud collisions and local gravitational
instabilities. This merging phase is where the dissipation occurs, and it tends to
maintain circular motions in the gas. The velocity components of the merged
particle are computed as mass weighted averages of the components of the
constituent particles. In most runs, the cell sizes are 0.1 units in radius and
$4^\circ$ in azimuth. Particles farther than 2.0 units from the
mid-plane of the disk are not merged, but continue on ballistic trajectories.
Since the disks remain relatively thin, this constraint has little effect other
than contributing to the disk thickness. With these cell sizes, the particles can
sometimes merge into relatively massive clouds containing several tens of mass
units, especially if the randomness of the explosions allows them to survive for
several explosion phases. Eventually each cloud is broken up in some explosion
phase, depending on the probabilities, and after redistribution of their
constituent particles through ballistic trajectories in the halo, other clouds
take their place.

Angular momentum and energy are explicitly conserved in
the ballistic phase, i.e., particles that scatter outward and fall back at
larger radii will have lower azimuthal speeds than those they meet in the
disk.

\section{Results and Discussion}
\label{results}

The algorithm of the previous section was run with a wide range of parameters to
determine the evolution of surface density profiles in different regimes (see below). An example is shown in Figure 1 where the disk begins with a flat radial profile and
evolves to an exponential. The halo potential corresponds to a slowly rising
rotation curve typical of a low mass galaxy, $v_{cir} \sim r^{0.3}$
\citep{persic96}. The upper panels show the particle distribution at the end of
the run ($100\ time\ units$), and after the feedback/exploding and ballistic travel phases. After the merging phase all particles would be located at the centers of adopted
grid elements. The particle distribution is smooth and the disk appears moderately
thick. A large majority of the particles lie in a quite thin disk.  A negligible fraction ($8/14040 = 0.057\%$) of the material has
been pushed to more than $2$ spatial units out of the disk, where it temporarily
does not participate in the merging and exploding phases.

The panel in the lower left shows the surface density profile at the initial and final times. The latter is close to an exponential form.

The lower right panel shows the initial rotation curve as a solid line, the final azimuthal
rotation speed as open triangles, and the radial and vertical ($z$) velocity
dispersions as asterisks and circles. The azimuthal velocity lags behind the
initial circular rotation curve because of the random motions. The ratio of the
velocity dispersion to the azimuthal velocity is substantially lower with this
cloud scattering than it is in models with stellar scattering off interstellar
clumps and holes (Papers I, II). The dissipation in the merging phase is
responsible. The final time shown corresponds to about $2.0$ Gyr in this dwarf
scaling, and the exponential starts to appear in about half of that time.

In the model of Fig. 1, the average magnitude of the velocity impulse given to
each fragment in each coordinate direction is 1.5 units, or a total average
magnitude of $1.5 \times 3^{0.5}$ units. This value equals $57$ km s$^{-1}$ in
the dwarf scaling given above and is reasonable or even conservative for shell and supershell
ejections off the plane. For example, \citet{relano07} find expansion velocities
of $50-100$ km s$^{-1}$ in the H$\alpha$ shells around OB associations in
local galaxies. 

Fig. 2 shows a case with a flat rotation curve and a small average relative
feedback velocity of 0.45 units, corresponding to an explosion velocity of $39$
km s$^{-1}$ in the disk scaling above. The figure shows that this small feedback yields a very thin disk, and an average dispersion to azimuthal velocity ratio of less than a tenth. The surface density profile in this case is a Type II form by the final time, which corresponds to about $4.0$ Gyr in the disk scaling. Slow explosion velocities take longer to form the exponential.

The half-mass height in Fig. 1, the thicker case, is 0.13 units. Taking the radius to be about 10 units we get a ratio of height to radius equal to 0.013. This is smaller than in NGC 891, a vigorously star-forming, edge-on disk galaxy.
According to \cite{bocchio16}, the dust scale height of N891 is $1.44\pm0.12$ kpc,
and the galaxy radius in NED\footnote{NASA/IPAC Extragalactic Database} is 6.75
arcmin, which is 17.8 kpc with the scaling of  2.64 kpc per arcmin.  Thus the
ratio of the dust scale height to radius in NGC 891 is $0.081$, a factor of 6
larger than in the figure. In the case shown in Fig. 2, the half-mass height to radius ratio is 0.0022, much smaller than in Fig. 1, as we might expect in this very mild case. 

Fig. 3 is for a case with a flat rotation curve and a larger average relative
feedback velocity of 0.75 units, corresponding to an explosion velocity of $65$
km s$^{-1}$ in the disk scaling above. In addition, this model also had the merger grid cell sizes decreased by a factor of two in both radial and azimuthal directions. The particle number was also increased by more than a factor of four (to 58,560), so that the clouds are more like giant clouds in present-day disks than very massive clouds in young disks. However, this change had little effect on the profile evolution when compared to a comparable model with the original grid and particle number. The radial profile evolves more rapidly in this case, and generates larger velocity dispersions and a somewhat thicker disk than in Fig. 2. The exponential form starts to appear at a time of a little over $1.0$ Gyr. By the late time shown in the figure the profile has nearly a single exponential form.
Despite the larger explosion velocity input (using the disk scaling), this flat rotation curve model retains a thin disk. Indeed, compared to recent cosmological galaxy formation models, the feedback seems quite modest (e.g., \citealt{christensen16},  \citealt{hopkins18}). In our models increased feedback magnitude leads to thicker disks, but this is not necessarily the case in self-consistent cooling models.

We have also done another run like that in Fig. 3, but with the initial particle mass further reduced by a factor of 5, the cell size reduced by 10\% and the feedback magnitude increased by 20\%. The results are essentially the same as Fig. 3. These results emphasize a couple of points: 1) a slightly increased feedback magnitude can significantly decrease the profile change timescale, and 2) smaller, but more numerous star-forming clouds can have the same effect as a smaller population of more massive clouds.

In sum, these models show that fountain feedback scattering
can generate exponential or S\'ersic type density profiles, with S\'ersic index close
to 1.0 in the latter case. We have carried out many more runs with different model
parameter values, and found that this is a general result as long as
the explosive input velocities of the feedback phase are sufficient to propel some
gas elements over significant radial excursions, though generally over a modest fraction of their initial radii (see upper right panel in Fig. 3 for examples).  Specifically, we have carried out models with: 1) rotation curves ranging from nearly Keplerian to nearly solid body, 2) explosive impulse velocities ranging over an order of magnitude, 3) ballistic phase timescales ranging over a factor of a few, and 4) particle numbers and cell sizes ranging over a factor of a few. All of these models tend toward exponential profiles, albeit on timescales that depend on the parameter values. It appears that this is essentially a diffusion process, which is not sensitive to the details of individual scattering events.
For reasonable values of the feedback velocity input these gas disks can also be quite cold.

\section{Conclusions}
\label{concl}

As discussed in the Introduction, exponential disks have been a persistent
mystery. Their universality suggests a common, robust generation mechanism
associated with galaxy formation, but disruptions from global disturbances like
interactions and accretions also suggest a second mechanism involving disk
restoration that works relatively rapidly. In a series of papers, we have
demonstrated that stellar scattering off massive clumps and holes in the
interstellar medium can generate stellar exponential profiles on timescales
typically less than a Gyr.  However, these stellar processes generally yield thick
disks, not cold, thin exponentials. Radial migration from spirals \citep{se02}
could, in principle, form an exponential-like profile while maintaining
near-circular orbits, but recent work by \citet{daniel18} suggests that too few
stars would migrate over large enough radial distances to rearrange the density
profile on the required timescale.

In the previous sections we have shown how scattering
of clouds in supershells and local fountains can drive the gas
towards an exponential profile in about a Gyr.
Profile change is a diffusive process averaged over many such events. While star-clump scattering tends to produce thick stellar disks,
gas scattering alone is dissipative and can produce thin disks.
Depending on parameter values, it can do so on short timescales, facilitating the reformation of disturbed exponential thin disks. 

In addition there is an evolutionary sequence in both the models above and the previous stellar scattering models (Papers I, II). Initially scattering processes, in flat-to-moderately rising rotation curve potentials, throw particles to larger radii, even to twice the initial radius with strong scattering. This extended disk usually has a steep exponential (Type II) profile. The evolution of the rest of the initial disk to an exponential profile is much slower, so a shallow exponential is present there for a long time (see Fig. 2). In the models here the break in the slope between the two exponentials marks the initial outer radius. Eventually, the scattering works to produce a single exponential (Fig. 3).

These models suggest a two-part solution to disk maintenance. Star-forming
disks generate exponential gas profiles through cloud scattering by feedback while
stars formed in these clouds inherit the profiles. Plus, stellar scattering by
clumps and holes operates at the same time, also forming an exponential, but
without needing to do all of the rearrangement by itself. In this way, the
velocity dispersion of the stars stays reasonably low and the disk maintains an
exponential on sub-Gyr timescales.

\section{Acknowledgements}

We are grateful to an anonymous referee for helpful comments that significantly clarified the presentation.

\newpage
\begin{figure*}
\epsscale{1.}
\plotone{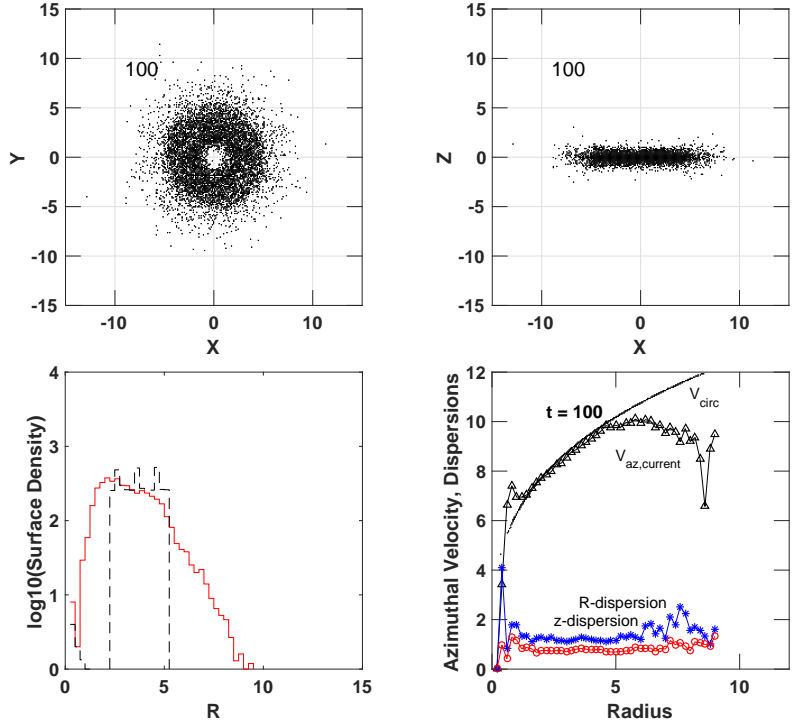}
\caption{The top two panels show the cloud particle distribution in two projections
at the end of a run in a model with a fixed rising rotation curve halo potential.
Particle positions are shown after the feedback/exploding and ballistic travel phases.
All panels are in dimensionless units; see text for representative scalings. The lower
left panel show the binned density profiles at the beginning of the run (black histogram),
and at the end ($t = 100$ units, red histogram). The lower right panel shows velocity
profiles, including: the circular velocity (black curve), the current azimuthal
velocity profile (triangles), the radial velocity dispersion (blue stars), and the
vertical velocity dispersion (red circles).} \label{fig1}
\end{figure*}

\newpage
\begin{figure*}
\epsscale{1.}
\plotone{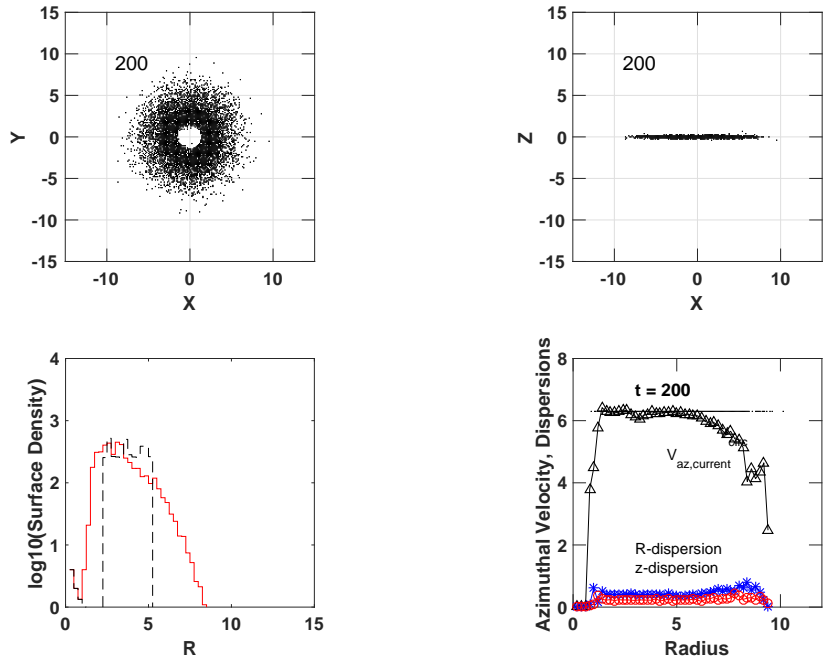}
\caption{Same as Fig. 1, but for a flat rotation curve potential, and with weaker feedback (see text). These models are
initialized with a central hole as another means to monitor scattering effects. The disk
also remains thinner, and the ratio of azimuthal velocity to velocity dispersions
is higher than in the model of Fig. 1.} \label{fig2}
\end{figure*}

\newpage
\begin{figure*}
\epsscale{1.}
\plotone{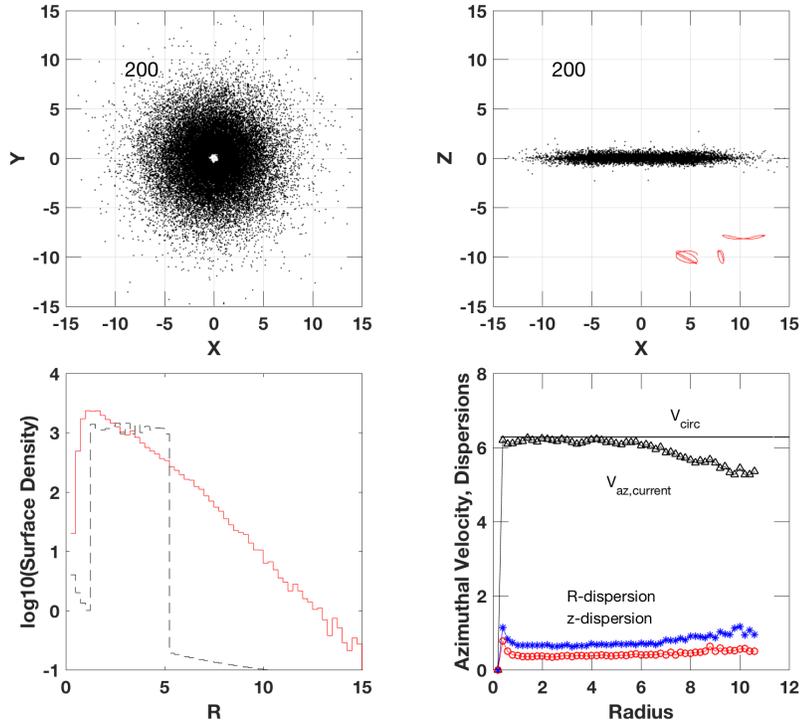}
\caption{Like Fig. 2, with a flat rotation curve potential, but here with stronger  feedback. Also the cell size of the merger grid is reduced by a factor of four and the particle number increased by about the same factor. The upper right panel shows three sample $r-z$ trajectories from the final ballistic phase of the run. These have been displaced by $8$ or $10$ units in $z$, and enlarged by a factor of $4$ in both $r$ and $z$ coordinates for visibility. } \label{fig3}
\end{figure*}

\end{document}